\documentclass[
preprint,
amsmath,amssymb,
aps,prd,
]{revtex4-1}

\usepackage[sort&compress]{natbib}
\usepackage{wrapfig}
\usepackage{dblfnote}
\DFNalwaysdouble
\usepackage{slashed}
\usepackage{graphicx}
\usepackage{dcolumn}
\usepackage{bm}
\usepackage{hyperref}

\begin{document}

\begin{flushright}
CERN-TH-2019-192
\end{flushright}

\title{Primordial black hole mergers, gravitational waves\\ and scaling relations}

\author{Alexandre Arbey}
 \altaffiliation[Also at ]{Institut Universitaire de France, 103 boulevard Saint-Michel, 75005 Paris, France}%
 \email{alexandre.arbey@ens-lyon.fr}
 \affiliation{Univ Lyon, Univ Claude Bernard Lyon 1, CNRS/IN2P3, Institut de Physique des 2 Infinis de Lyon, UMR 5822, F-69622, Villeurbanne, France}
 \affiliation{Theoretical Physics Department, CERN, CH-1211 Geneva 23, Switzerland}

\author{Jean-Fran\c{c}ois Coupechoux}
 \email{j-f.coupechoux@ipnl.in2p3.fr}

\affiliation{Univ Lyon, Univ Claude Bernard Lyon 1, CNRS/IN2P3, Institut de Physique des 2 Infinis de Lyon, UMR 5822, F-69622, Villeurbanne, France}

\begin{abstract}

Observations of gravitational waves provide new opportunities to study our Universe. In particular, mergers of stellar black holes are the main targets of the current gravitational wave experiments. In order to make accurate predictions, it is however necessary to simulate the mergers in numerical general relativity, which requires high performance computing. While scaling relations are used to rescale simulations for very massive black holes, primordial black holes have specific properties which can invalidate the rescaling. Similarly black holes in theories beyond Einstein's relativity can have different scaling properties. In this article, we consider scaling relations for the most general cases of primordial black holes, such as charged and spinned black holes, and study the effects of the cosmological expansion and of Hawking evaporation. We also consider more exotic black hole models and derive the corresponding scaling relations, which can be compared to the observations in order to identify the underlying black hole model and can be used to rescale the numerical simulations of exotic black hole mergers.

\end{abstract}

\pacs{Valid PACS appear here}
\maketitle


\section{Introduction}
\label{sec:intro}

After the discovery of gravitational waves (GWs) by LIGO \cite{Abbott:2016blz}, studies of mergers of stellar mass black holes (BHs) and of the resulting emission of gravitational waves have multiplied, either with semi-analytical descriptions or with numerical general relativity simulations. In absence of discovery of new particles at the LHC and in dark matter detection experiments, the nature of dark matter is still actively searched for, and the fact that primordial black holes (PBHs) can constitute dark matter is now considered as a viable possibility \cite{Carr:2009jm,Carr:2016drx,Carr:2020xqk}. Contrary to stellar black holes, the mass of primordial black holes spans from values as low as the Planck mass up to millions of solar masses, and the merger of such PBHs can generate GWs with frequencies and amplitudes very different from those accessible to LIGO and VIRGO. Yet such PBH-generated GWs may be accessible to future GW experiments such as eLISA \cite{AmaroSeoane:2012km}, and it is important to correctly model them. However because the masses of the involved PBHs can be very different from the stellar BHs for which numerical simulation codes have been built, it may be very difficult to simulate numerically their mergers, and in particular the final states, which requires numerical simulations for a correct description. Fortunately there exists scaling relations which can in principle be used to rescale results obtained in numerical simulations of stellar BH mergers to very different mass scales and which also apply to GW emission. Nevertheless, some particularities of the primordial black holes, such as very small masses, maximal spins, non-negligible charges and Hawking evaporation, can falsify such scaling relations. We study in this article the scaling properties of primordial black holes, and determine the limits of validity of the scaling relations. We also consider different cases of non-standard black hole models and study their scaling properties. The modified scaling properties can then be used to distinguish between black hole models, and to accelerate the generation of catalogues of simulated non-standard black hole mergers. In Section 2 we explicitly retrieve the scaling relations of standard black holes and derive scaling relations for the most general case, i.e. Kerr-Newman black holes with spin and electromagnetic charge. In Section 3 we discuss the domain of validity of such a rescaling, taking into account Hawking evaporation and expansion of the Universe. In Section 4, we study whether non-standard models of black holes can also have specific scaling properties. In Section 5, we consider the limitations of numerical simulations of black hole mergers in the case where the standard scaling properties are not valid and discuss the advantages of identifying new scaling relations, before concluding. 

In the following, we use the natural unit system with $c \equiv 1$.

\section{Scale invariance of standard black holes}

\subsection{Scale invariance}
\label{sec:scaling}

In the coordinate system $x^\mu = (t,x,y,z)$ associated to a metric $g_{\mu\nu}$, we consider a transformation with a scale factor $\lambda$ such as
\begin{equation}
 x^\mu \longrightarrow \lambda x^\mu \,. \label{eq:transformation}
\end{equation}
By definition of the metric and its relation with the local flat coordinates, the metric is scale invariant under transformation~(\ref{eq:transformation}). Under this transformation, we obtain for the proper time $\tau$, Christoffel symbol $\Gamma^\sigma_{\mu\nu}$ and Riemann tensor $R^\sigma_{\;\;\mu\nu\kappa}$:
\begin{eqnarray}
 d\tau^2 &=& g_{\mu\nu} dx^\mu dx^\nu \longrightarrow \lambda^2 d\tau^2\,,\nonumber\\
 \Gamma^\sigma_{\mu\nu} &=& \frac12 g^{\sigma\alpha}\left( \frac{\partial g_{\alpha\mu}}{\partial x^\nu} + \frac{\partial g_{\alpha\nu}}{\partial x^\mu} - \frac{\partial g_{\mu\nu}}{\partial x^\alpha} \right) \longrightarrow \lambda^{-1} \Gamma^\sigma_{\mu\nu}\,,\\
 R^\sigma_{\;\;\mu\nu\kappa}&=&\frac{\partial \Gamma^\sigma_{\mu\kappa}}{\partial x^\nu}-\frac{\partial \Gamma^\sigma_{\mu\nu}}{\partial x^\kappa} + \Gamma^\sigma_{\mu\alpha}\Gamma^\alpha_{\mu\kappa} - \Gamma^\sigma_{\kappa\alpha}\Gamma^\alpha_{\mu\nu} \longrightarrow \lambda^{-2} R^\sigma_{\;\;\mu\nu\kappa}\nonumber
\end{eqnarray}
In addition the four velocity vector $u^\mu = dx^\mu/d\tau$ is scale invariant, and the Ricci tensor $R_{\mu\nu}$, the scalar curvature $R$ and the Weyl conformal tensor $C_{\sigma\mu\nu\kappa}$ scale similarly to the Riemann tensor $R^\sigma_{\;\;\mu\nu\kappa}$. The geodesic equation 
\begin{equation}
 \frac{d^2 x^\mu}{d\tau^2} + \Gamma^\mu_{\alpha\beta} \frac{dx^\alpha}{d\tau} \frac{dx^\beta}{d\tau} = 0
\end{equation}
is also scale invariant, and therefore its solutions are unchanged and only the coordinates are dilated.

Concerning Einstein's field equations (EFEs) with a cosmological constant $\Lambda$ and a stress-energy tensor $T_{\mu\nu}$:
\begin{equation}
 R_{\mu\nu} - \frac12 g_{\mu\nu}R + \Lambda g_{\mu\nu} = 8\pi G \, T_{\mu\nu} \,,
\end{equation}
it is clear that the EFEs are not scale invariant. Indeed the cosmological constant is not scale invariant, neither is $T_{\mu\nu}$ in the general case. The scale invariance is retrieved in absence of cosmological constant and in the vacuum.

The description of the merger of black holes and generation of gravitational waves fulfils the scale-invariance condition: the GWs are perturbations of the metric in the vacuum, and the BHs are described as metrics with horizons. In absence of matter outside the BHs, the EFEs are written in the vacuum, and the cosmological constant is too small to have any effect during the black hole coalescence and merger. Let us consider the Kerr metric \cite{Kerr:1963ud} in the Boyer–Lindquist coordinates \cite{Boyer:1966qh}, which describes a rotating black hole of mass $M$ and angular momentum $J$:
\begin{equation}
 d\tau^2 = \big(dt - a \sin^2\theta d\phi\big)^2\, \frac{\Delta}{\Sigma} - \left(\frac{dr^2}{\Delta} + d\theta^2\right) \Sigma - \big((r^2+a^2)d\phi-a dt\big)^2 \, \frac{\sin^2\theta}{\Sigma} \,,\label{eq:kerr}
\end{equation}
where ($r,\theta,\phi$) are spherical coordinates, $a=J/M$, $\Sigma=r^2 + a^2 \cos^2\theta$, $\Delta = r^2 - R_s r + a^2$ and $R_s=2 GM$ the Schwarzschild radius. In the case of vanishing angular momentum, the Schwarzschild metric is retrieved:
\begin{equation}
 d\tau^2 = \left(1 - \frac{2 GM}{r}\right) dt^2 - \left(1 - \frac{2 GM}{r}\right)^{-1} dr^2 + r^2 (d\theta^2 + \sin^2\theta d\phi^2)\,.
\end{equation}

The scale invariance of the Kerr metric implies that:
\begin{eqnarray}
 R_s \longrightarrow \lambda R_s &\Longleftrightarrow& M \longrightarrow \lambda M \,, \label{eq:mass_transformation}\\
 a \longrightarrow \lambda a\;\; &\Longleftrightarrow& \;\;J \longrightarrow \lambda^2 J \,,
\end{eqnarray}
under transformation~(\ref{eq:transformation}). The transformation of $J$ is compatible with the standard definition of angular momentum $\vec{J}=m \,\vec{x} \times \vec{v}$ when applying the coordinate and mass transformations (\ref{eq:transformation}) and (\ref{eq:mass_transformation}). Also the Kerr dimensionless spin parameter $a^* = a/M$, which is equal to 0 for Schwarzschild BHs and 1 for extremal Kerr BHs, is scale invariant.
Regarding the mass transformation, a remark is in order: for a BH the mass is not a real physical mass but a measure of the horizon radius. 

We can now derive the following scaling rules for the merger of $n$ BHs of masses $M_i$, spins $J_i$ and momentum $\vec{P}_i$ with positions $\vec{x}_i$ ($i=1\cdots n$) at time $t$:
\begin{eqnarray}
 M_i \longrightarrow \lambda M_i &\;,\hspace*{1.cm}& t \longrightarrow \lambda t\,, \\
  \vec{P}_i \longrightarrow \lambda \vec{P}_i &\;,\hspace*{1.cm}& \vec{x}_i \longrightarrow \lambda \vec{x}_i\,, \nonumber \\
 J_i \longrightarrow \lambda^2 J_i &\;,\hspace*{1.cm}& a^*_i \longrightarrow a^*_i\,. \nonumber
\end{eqnarray}
Consequently the local densities scale as $\rho \longrightarrow \lambda^{-2} \rho$, the accelerations as $\vec{a} \longrightarrow \lambda^{-1} \vec{a}$ and the velocities are scale invariant.

Similarly, since gravitational waves can be considered as perturbations of the metric, their frequency $f$, wavelength $\Lambda$, energy $E$, amplitude (or metric perturbation) $h$, speed $v$ and stress-energy tensor scale as
\begin{eqnarray}
 f \longrightarrow \lambda^{-1} f &\;,\hspace*{1.cm}& \Lambda \longrightarrow \lambda \Lambda\;, \label{eq:gw_rules}\\
 E \longrightarrow \lambda E \;\;\;&\;,\hspace*{1.cm}& h \longrightarrow h\;, \nonumber\\
 v \longrightarrow v \hspace*{0.6cm}&\;,\hspace*{1.cm}& T^{\mu\nu} \longrightarrow \lambda^{-2} T^{\mu\nu} \;. \nonumber
\end{eqnarray}

\subsection{Scaling relations of charged spinning black holes}
\label{sec:kerr_newman}

While stellar black holes are generally expected to have small spins and negligible electromagnetic charges, primordial black holes can have very small masses, extremal spins and be strongly charged. In particular, the lifetime of primordial black holes can be strongly increased by an electromagnetic charge \cite{Sorkin:2001hf}, making possible the existence of the very light ones even today.

In the general case, primordial BHs have to be described by the Kerr-Newman metric \cite{Newman:1965my}, which is similar to Eq.~(\ref{eq:kerr}), with $\Sigma=r^2-R_s r + a^2 + R_Q^2$, where $R_Q$ is related to the black hole charge Q by
\begin{equation}
 R_Q = \sqrt{\frac{G}{4\pi \epsilon_0}} Q \,,
\end{equation}
with $\epsilon_0$  the void permittivity. The scale invariance can be restored by transforming the charge as
\begin{equation}
 Q \longrightarrow \lambda Q\,.
\end{equation}
However, such a rotating charge induces an electromagnetic potential such as
\begin{equation}
 A_\mu = \left( \frac{r R_Q}{\Sigma}, 0 , 0 , - \frac{a^* R_Q r \sin^2\theta}{\Sigma G}\right)\,,
\end{equation}
which is invariant under our set of transformations. The electromagnetic field therefore transforms as
\begin{equation}
 F_{\mu\nu}=\frac{\partial A_\nu}{dx^\mu} - \frac{\partial A_\mu}{dx^\nu}
\longrightarrow \lambda^{-1} F_{\mu\nu}\,,
\end{equation}
and the associated stress-energy tensor as
\begin{equation}
 T^{\mu\nu}=\epsilon_0 \left(F^{\mu\alpha} g_{\alpha\beta} F^{\nu\beta} - \frac14 g^{\mu\nu} F_{\delta\gamma} F^{\delta\gamma} \right) \longrightarrow \lambda^{-2} T^{\mu\nu} \,,
\end{equation}
which is the behaviour required to let the EFEs invariant in presence of a source term. Turning to Maxwell's equations:
\begin{equation}
 \frac{\partial }{dx^\nu}\left(\sqrt{-\det(g_{\mu\nu})} g^{\mu\alpha} F_{\alpha\beta} g^{\beta\nu} \right) = \frac{J^\mu}{\epsilon_0} \,,
\end{equation}
they are invariant when the current four-vector $J^\mu$ is zero or scales as $\lambda^{-2}$. Therefore, not only a Kerr-Newman BH respects the scaling relation described in Section~\ref{sec:scaling} and charge normalized to the BH mass $Q/M$ is scale invariant, but the electromagnetic waves emitted by a merger of charged black holes in empty space respect the same scaling properties as gravitational waves, given in Eq.~(\ref{eq:gw_rules}). There is indeed a strong parallel between Einstein's field equations and Maxwell's equations: in presence of source terms the scaling properties are generally broken.

As a consequence, any simulation of charged spinning black holes can be rescaled to larger or smaller masses, and the gravitational waves and the electromagnetic waves can be rescaled similarly.

\section{Domain of validity of the scaling relations for primordial black holes}
\label{sec:limitations}

Primordial black holes are by definition created in the primordial Universe, and can have masses as low as the Planck mass. With such small masses, quantum effects have to be taken into account, and interaction of the horizon of light black holes with the vacuum results in the BH evaporation by emission of particles, called Hawking radiation \cite{Hawking:1974sw}. In the following, we study the domain of validity of the scaling relation with respect to surrounding material acting as source terms, expansion of the Universe and evaporation of BHs.

\subsection{Surrounding material around black holes}

Let us briefly consider the presence of source terms in the EFEs. For a perfect fluid in thermodynamic equilibrium, the stress-energy tensor reads:
\begin{equation}
 T^{\mu\nu} = (\rho + P) u^\mu u^\nu + P g^{\mu\nu} \,,
\end{equation}
where $\rho$ and $P$ are the density and pressure of the fluid. In order to leave the EFEs scale invariant, the density and pressure need to transform under the scale transformation as:
\begin{equation}
 (\rho, P) \longrightarrow \lambda^{-2} (\rho, P)\,. \label{eq:rhoP}
\end{equation}
Such a transformation does not hold in the general case, but it is valid for collisionless particles with negligible pressure, or when $P \propto \rho$, or when both pressure and density are negligibly small. This has been studied numerically for example in \cite{Giacomazzo:2012iv,Kelly:2017xck}.

In the general case, when processes dominated by electromagnetic, weak or strong interactions occur the scaling is broken, because these interactions are related to different couplings, and in addition weak and strong interactions have limited ranges. Similarly, quantum effects cannot be expected to be scale-invariant since they are independent from gravity.

\subsection{Black holes and expansion rate of the Universe}

\begin{figure}[!t]
\begin{center}
\includegraphics[width=0.8\linewidth, trim = 0cm 0cm 5.7cm 0cm, clip]{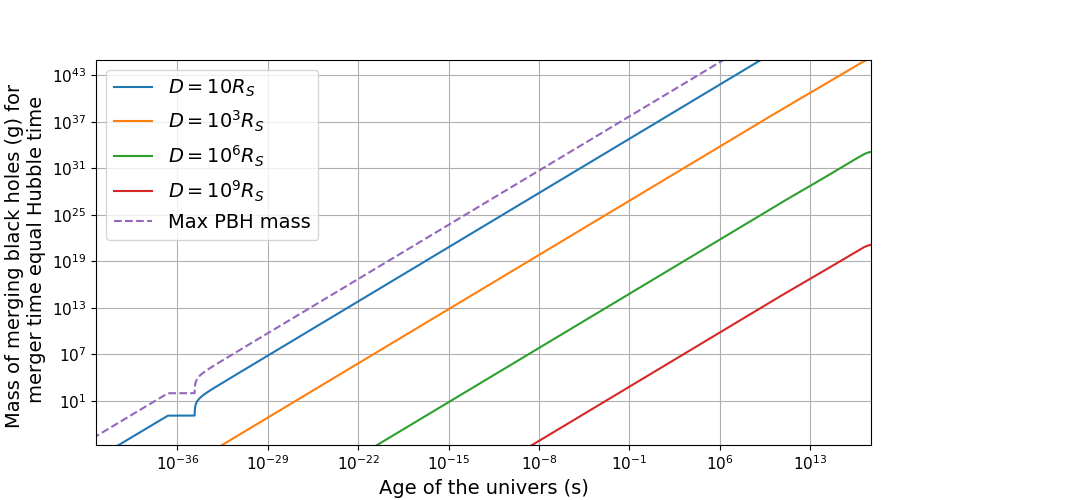}
\end{center}
\caption{BH masses corresponding to a merger time equal to the Hubble time, as a function of the age of the Universe, for different initial merger BH distances $D$ given as numbers of Schwarzschild radii. The dashed line corresponds to the maximum mass of BHs.\label{fig:cosmo}}
\end{figure}

Since PBHs originate in the early Universe, their mergers can be affected by the expansion. As the expansion is time-dependent but affects space, it breaks the scaling of time and space. A simple rule would be to consider that a rescaling is possible as long as the duration of the merger $T_{\rm merger}$ is much smaller than the Hubble time $t_{\rm Hubble}(t)$ at cosmological time $t$, that is
\begin{equation}
T_{\rm merger} \ll t_{\rm Hubble}(t) \left[ \equiv \frac{a(t)}{\dot{a}(t)} \right] \,,
\end{equation}
where $a(t)$ is the cosmological scale factor. To calculate the merger time, we approximated the trajectory of the BHs to an equilibrium circular orbit with an orbital decay rate $dD(t)/dt$. At lowest order, this expression can be calculated from the quadrupole formula \cite{Baumgarte:2010ndz} and an integration of the orbital decay gives, for the binary separation distance $D(t)$ of the coalescence of two Schwarzschild BHs of Schwarzschild radius $R_s$, the following scale invariant expression:
\begin{equation}
D(t)=4\left[\frac{R_s^3}{20}(T_{\rm merger}-t)\right]^{1/4}\,,\label{eq:distance}
\end{equation}
which is valid at $t<T_{\rm merger}$. Assuming a $\Lambda$CDM model in a flat Universe, using Planck 2018 cosmological parameters~\cite{Aghanim:2018eyx} and considering 60 e-folds for inflation, we show in Fig.~\ref{fig:cosmo} the individual BH masses of binary mergers for which $T_{\rm merger} = t_{\rm Hubble}(t)$, as a function of the age of the Universe $t$. BHs with masses above the lines merge faster than the expansion. The line corresponding to the maximal PBH mass has been obtained by assuming that the biggest BH has a Schwarzschild radius equal to the Hubble radius. In particular, at our present epoch, mergers of stellar black holes are affected by the expansion only if their initial distance is larger than about~$10^6\,$km. 

\subsection{Black holes and Hawking evaporation}

\begin{figure}[!t]
\begin{center}
\includegraphics[width=0.8\linewidth, trim = 0cm 0cm 4.9cm 0cm, clip]{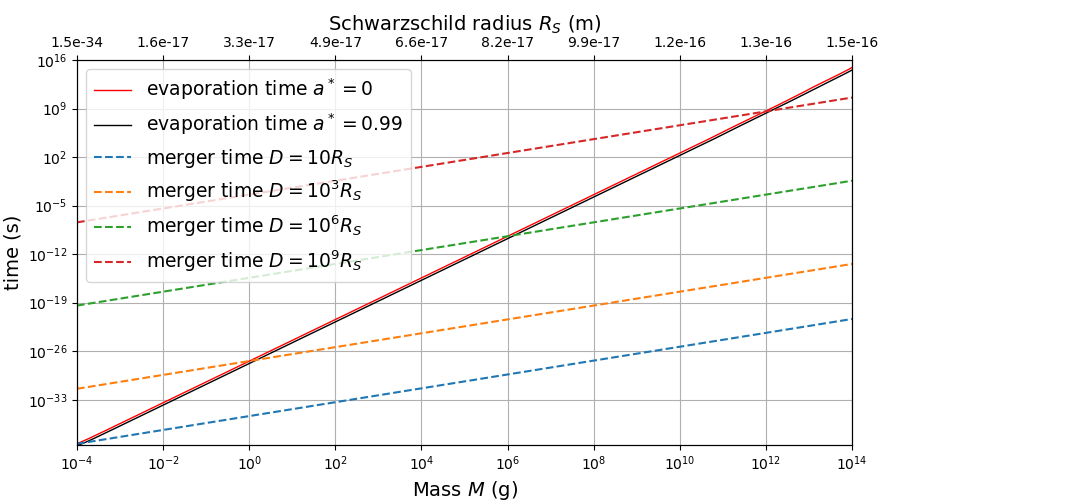}
\end{center}
\caption{In the time vs. BH mass plane, the solid lines correspond to the evaporation time for a BH of mass $M$ and spin $a^*=0$ (red) and $a^*=0.99$ (blue). The dashed lines correspond to the merger time for two identical Schwarzschild BHs of mass $M$, for different initial distances $D$ given as numbers of the Schwarzschild radius.\label{fig:hawking}}
\end{figure}

Light PBHs are expected to vanish via emission of Hawking radiation~\cite{Hawking:1974sw}. Since the lifetime of a BH is typically proportional to its mass cubed~\cite{Page:1976df}, the scaling of the mass proportionally to the spacetime scaling is not possible anymore. Therefore, the scaling can be applied only if the duration of the merger is much smaller than the lifetime of the BHs. In Fig.~\ref{fig:hawking} we show the evaporation time of BHs as a function of their mass, for Schwarzschild BHs and for nearly extremal Kerr BHs with $a^*=0.99$. The evaporation time has been computed with the public program {\tt BlackHawk}~\cite{Arbey:2019mbc}. For comparison, typical durations of mergers of two black holes obtained from Eq.~(\ref{eq:distance}) are also plotted as a function of the mass of the system, assuming that both BHs have identical masses, for initial distances $D =(10\, R_s,10^3\, R_s, 10^6\, R_s, 10^9\, R_s)$, so that the rescaling is typically correct down to masses of about $(10^{-4}, 1, 10^6, 10^{12})$ grams, respectively, below which evaporation has to be taken into account and breaks the rescaling. Since PBHs having not completely evaporated today have masses above $10^{14}\,$g \cite{Arbey:2019jmj,Arbey:2019vqx}, the mergers of the surviving PBHs are not affected by Hawking evaporation.

\section{Non-standard black holes}

In this section, we study the scaling properties of non-standard black holes, which can either be extensions of the Schwarzschild black hole model within Einstein-gravity, or black holes in non-standard gravity theories. In the latter case, not only the generation of gravitational waves during a black hole merger can be modified, but the propagation of the gravitational waves itself can be affected. Nevertheless, since the propagation of gravitational waves mostly occurs in a weak gravitational field, that the observed gravitational waves are in agreement with general relativity, we will disregard the modification of propagation in weak gravitational fields and mainly discuss the scaling properties of non-standard black holes and gravity scenarios.

\subsection{Black holes in expanding Universe}

We first consider the McVittie metric \cite{McVittie:1933zz}, which describes a non-spinning and neutral black hole in a flat expanding Universe:
\begin{equation}
 d\tau^2 = \left(\frac{1-\mu}{1+\mu}\right)^2 dt^2 - (1+\mu)^4 a^2(t)(d\rho^2+\rho^2\, d\Omega^2)\,,
\end{equation}
where $\rho$ is the comoving radius, $a(t$) the expansion factor, $d\Omega^2=d\theta^2 + \sin^2\theta d\phi^2$ is the solid angle and $\mu=GM/[2a(t)\rho]$. This metric coincides with the Schwarzschild metric for a constant $a(t)$ under the redefinition $r=a\,\rho$, and with the Robertson and Walker metric in flat space at large radius.
The scaling relations are preserved as long as $a(t) \rho \longrightarrow \lambda\,a(t) \rho$. In the case of a constant scale factor, the scaling relation $\rho \longrightarrow \lambda\rho$ is similar to transformation~(\ref{eq:transformation}). However, in presence of expansion, it is more interesting to consider that the scaling relation would read instead $a(t) \longrightarrow \lambda\,a(t)$. In a radiation-dominated Universe $a(t) = (2 \pi G\rho_0^r/3)^{1/4}  t^{1/2}$, and in a matter-dominated Universe $a(t) =(6 \pi G\rho_0^m)^{1/3} t^{2/3}$, where $\rho_0^r$ and $\rho_0^m$ are the current radiation and matter densities, respectively. Therefore, as long as the dominating component is not changing, the scaling relations are preserved when $\rho_0^r \longrightarrow \lambda^2\,\rho_0^r$ or $\rho_0^m \longrightarrow \lambda\,\rho_0^m$. This result is confirmed when considering that an expanding Universe necessarily involves a non-zero stress-energy tensor, and the density and pressure have the scaling properties given in Eq.~(\ref{eq:rhoP}). Using the transformations of $\rho_0^m$ and $\rho_0^r$, we check that for matter $\rho^m = \rho_0^m a^{-3} \longrightarrow \lambda^{-2} \rho^m$, $P^m \approx 0$, and for radiation $\rho^r = \rho_0^r a^{-4} \longrightarrow \lambda^{-2} \rho^r$ and $P^r = \rho^r / 3$, so that the needed scaling properties are respected by the stress-energy tensor. Thus, simulations of black hole mergers in expanding Universe can be rescaled as long as the dominating component is conserved and $\rho_0^r$ or $\rho_0^m$ are rescaled properly. However, it is also clear that there is no possibility to use a rescaling transformation from a static Universe to an expanding one.

\subsection{Morris-Thorne traversable wormholes}

We consider here the Morris-Thorne traversable wormhole metric \cite{Morris:1988cz}, which is spherically-symmetric and static:
\begin{equation}
 d\tau^2 = e^{2\Phi(r)} dt^2 - \frac{dr^2}{1-b(r)/r} + r^2(d\theta^2 + \sin^2\theta d\phi^2)\,, \label{eq:morris}
\end{equation}
where $\Phi(r)$ and $b(r)$ are arbitrary functions of $r = r_0 + l$, where $r_0$ is the radius of the wormhole throat and $l$ the distance from the throat. A wormhole is connected to two universes, so that a value $r > r_0$ corresponds to two points, one in our Universe and one on the other side of the wormhole. For such a wormhole to exist, the $\Phi(r)$ should not have any singularity and $1-b(r)/r$ has to remain positive for $r\le r_0$. Also at infinity $\Phi(r)$ and $b(r)$ go to zero.

Contrary to the case of black holes, there exists a non-zero diagonal stress-energy tensor~\cite{Lemos:2003jb} with elements:
\begin{align}
 T_{tt} &= \frac{1}{8\pi G}\frac{b'}{r^2}\,,\nonumber\\
 T_{rr} &= -\frac{1}{8\pi G}\left[ \frac{b}{r^3} -2 \left(1 - \frac{b}{r}\right) \frac{\Phi'}{r} \right]\,,\\
 T_{\theta\theta} = T_{\phi\phi} &= \frac{1}{8\pi G}\left[\Phi'' + \bigl(\Phi'\bigr)^2 - \frac{b' r - b}{2 r^2(1-b/r)}\Phi'- \frac{b' r - b}{2 r^3(1-b/r)} + \frac{\Phi'}{r} \right]\,.\nonumber
\end{align}

It is important to note that if the stress-energy tensor vanishes away from the throat, Birkhoff's theorem implies that the wormhole metric reduces to the Schwarzschild metric, so that an isolated wormhole is indistinguishable from the Schwarzschild black hole away from the center. The coalescence of two wormholes or a wormhole and a black hole would therefore be indistinguishable from the a pure binary black hole when they are far away, but the gravitational effects generated by the non-zero stress-energy tensor can affect the merger and the spectrum of emitted gravitational waves. Considering metric (\ref{eq:morris}), under the transformation of Eq.~(\ref{eq:transformation}), the scale invariance imposes that
\begin{eqnarray}
\Phi(r) \longrightarrow \Phi(r)\,,\\
b(r) \longrightarrow \lambda b(r)\,. \nonumber
\end{eqnarray}
These conditions automatically lead to $T_{\mu\nu} \longrightarrow \lambda^{-2} T_{\mu\nu}$ since wormholes are EFE solutions.

Looking at the zero tidal force hypothesis of Morris and Thorne \cite{Morris:1988cz}, which corresponds to $\Phi(r)=0$ and $b(r)=(b_0 r)^{1/2}$, the scale invariance implies that the constant $b_0$ scales as a distance.

Concerning the emission of gravitational waves during a merger involving two wormholes, the question of their scaling properties is therefore not related to the geometry, but to the exotic energy which feeds the stress-energy tensor, and in particular to the interaction of the exotic energy of one of the wormholes with the exotic energy of the other wormhole. If their interaction has similar scaling properties as gravity, then the emitted gravitational waves will have the same scaling properties as in a binary black hole merger, but it will not be the case otherwise.

\subsection{Loop quantum gravity inspired black holes}

One of the major theoretical questions about the Schwarzschild metric concerns the naked irreducible singularity at the center, which may find a solution in quantum gravity, in which the smallest quanta is expected to have a minimal size that could affect the properties of the black hole interior and in particular the geometry at its center. We consider here the loop quantum gravity (LQG) corrected Schwarzschild metric of \cite{Modesto:2008im}:
\begin{equation}
d\tau^2 = \frac{(r-r_+)(r-r_-)(r+r_*)^2}{r^4+a_0^2} dt^2 - \left(\frac{(r-r_+)(r-r_-)r^4}{(r+r_*)^2(r^4+a_0^2)}\right)^{-1} dr^2 - \left(r^2+\frac{a_0^2}{r^2}\right)d\Omega^2 \,,
\end{equation}
where $r_+=2 GM (1+P)^{-2}$, $r_-=2GM \,P (1+P)^{-2}$ and $r_*=\sqrt{r_+\,r_-}$, with $P=(\sqrt{1+\epsilon^2}-1)/(\sqrt{1+\epsilon^2}+1)$, $\epsilon$ being a very small constant, $a_0=A_{\rm min}/8\pi$ and $A_{\rm min}$ is the area corresponding to the smallest quanta in the LQG quantification. $r_-$ and $r_*$ are two small radii which prevent the singularity at $r=0$ in the Schwarzschild metric. The Schwarzschild metric is retrieved for $\epsilon=0$ and $A_{\rm min}=0$. For this metric, the scaling relations still hold as long as $A_{\rm min} \longrightarrow \lambda^2 A_{\rm min}$, which is consistent with the scaling of an area, and $\epsilon$ is scale-independent. However, it is clear that physically the smallest quanta area should not be rescaled, but the probable small size of $a_0$ would certainly not allow to distinguish the gravitational waves emitted during a merger of LQG corrected Schwarzschild black holes from the ones of a standard binary black hole merger.

\subsection{Black holes in $f(R)$ gravity}

We now consider the well-known generalization of the Einstein-Hilbert Lagrangian density, where $R$ is replaced by a generic function $f(R)$ \cite{Buchdahl:1983zz,Sotiriou:2008rp}. Such a modification encompasses a broad variety of new gravitational theories, including quantum gravity models. 

From the definition of the Lagrangian, in empty space the scaling relation $R \longrightarrow \lambda^{-2} R$ in Einstein's gravity becomes $f(R) \longrightarrow \lambda^{-2} f(\lambda^{-2} R)$, or equivalently $R \longrightarrow f^{-1}(\lambda^{-2} f(\lambda^{-2} R))$. This transformation does not obviously hold unless $f(R) = R$. It can nevertheless still be valid in weak gravitational fields at first order in $R$ when $f(R) \approx R$, therefore not affecting the propagation of gravitational waves.

The equation equivalent to EFEs reads in $f(R)$ gravity:
\begin{equation}
 f'(R) R_{\mu\nu} -\frac12 f(R) g_{\mu\nu} = [\nabla_\mu \nabla_\nu - g_{\mu\nu} \square]f'(R) + 8\pi G \,  T_{\mu\nu}\,, \label{eq:fR_motion}
\end{equation}
where $\nabla_\mu$ is the covariant derivative, $\square=\nabla^\mu \nabla_\mu$ is the Laplacian, and $T_{\mu\nu}$ is the standard stress-energy tensor. In the vacuum the right hand-side term appears as an effective stress-energy tensor. In case of a constant scalar curvature $R=R_0$ --- as for example in empty space in the Einstein's gravity --- the equation becomes in vacuum:
\begin{equation}
 f'(R) R_{\mu\nu} - \frac12 f(R) g_{\mu\nu} = 0\,,
\end{equation}
and its trace gives:
\begin{equation}
 f'(R_0) R_0 - 2 f(R_0) = 0\,,
\end{equation}
which admits as solution $f(R_0) \propto R_0^2$, or $R_0 = 0$ if $f(R_0)=0$. This second solution is the same of the one of general relativity in empty space, and the standard Schwarzschild metric is therefore a solution of the field equations (\ref{eq:fR_motion}) as long as $f(0)=0$ \footnote{The converse is not necessarily true, and other spherically-symmetric and static black hole metric solutions may also exist in $f(R)$ gravity, without being valid in Einstein's gravity.}. In such a case, the scaling properties of the Schwarzschild metric in $f(R)$ gravity are therefore the same as in Einstein's gravity.

A more careful look at the field equations (\ref{eq:fR_motion}) reveals that even in absence of the scaling $f(R) \longrightarrow \lambda^{-2} f(\lambda^{-2} R)$ the equation itself scales as $\lambda^{-2}$ in empty space, so that the scaling properties of general relativity in vacuum are conserved in $f(R)$ gravity. Therefore the merger of two Schwarzschild black holes in $f(R)$ gravity will have the same scaling properties as in Einstein's gravity, and the observed gravitational waves will also have the same properties.

Nevertheless, because of the different field equations, the gravitational waves emitted during the mergers and their propagation will be affected by $f(R)$ gravity. Thus, in the case of detection of non-standard gravitational waves, a scaling of the gravitational wave spectra similar to the one expected in Einstein's gravity would constitute a hint towards $f(R)$ gravity.

\subsection{Black holes in gravity with higher-curvature terms}

More generally, modified gravity can be based on Lagrangians with higher-curvature terms involving higher-rank tensors in addition to the scalar curvature as in $f(R)$ gravity, such as $R_{\mu\nu}R^{\mu\nu}$ or $R_{\alpha\beta\mu\nu}R^{\alpha\beta\mu\nu}$, or more complicated scalar objects such as
\begin{equation}
 (^*R^{\alpha\beta\mu\nu})\, R_{\alpha\beta\mu\nu} = \left(\frac12\frac{\epsilon^{\rho\sigma\mu\nu}}{\sqrt{-g}}R^{\alpha\beta}_{\;\;\;\;\rho\sigma}\right)\, R_{\alpha\beta\mu\nu}
\end{equation}
in Jackiw-Pi theory \cite{Jackiw:2003pm}, where $\epsilon^{\rho\sigma\mu\nu}$ is the totally antisymmetric tensor.

In general, as long $R_{\mu\nu}=0$ is a solution of the field equations, the Einstein's gravity Schwarzschild metric is a solution to describe static and spherically-symmetric black holes, and the scaling relations of general relativity in empty space will still be valid. This is the case for the three examples of scalars given above, but in general not for a cubic contraction of the Riemann tensor \cite{Deser:2003up}. If the Schwarzschild metric is valid, similarly to $f(R)$ gravity, the emitted gravitational waves and their propagation will be affected by the modified gravity, but the scaling properties will still hold.

\subsection{Extra-dimensional black holes}
\label{sec:xdim}

The generalization of general relativity to extra-dimensions is rather straightforward, and the geometrical part of the action in a $4+n$ dimensional spacetime reads:
\begin{equation}
 S_g = \frac{M^{2+n}_*}{16\pi} \int d^{4+n}x \sqrt{-g} (R_4 + R_n) \,,
\end{equation}
where $M_*$ is the fundamental mass scale, $R_4$ is the 4-dimensional part of the scalar curvature and $R_n$ is its $n$-dimensional part. An effective 4 dimension-action can be obtained by integrating over the extra-dimensions and by neglecting the $R_n$ term which only affects the geometry of the extra-dimensions, leading to:
\begin{equation}
 S_g^4= \frac{M^{2+n}_* L^n}{16\pi} \int d^4x \sqrt{-g} R_4\,,
\end{equation}
where it is assumed for simplicity that the extra-dimensions have the same size $L$. The Planck mass is therefore related to $M_*$ by
\begin{equation}
 M_P = M_*^{1+n/2} L^{n/2}\,.
\end{equation}
The number $n$ and the size $L$ of the extra-dimensions are therefore the main parameters of this model. The Schwarzschild solution for a spherically-symmetric and static metric also exists and reads \cite{Myers:1986un,Emparan:2008eg}:
\begin{equation}
 d\tau^2 = \left(1 - \left(\frac{r_h}{r}\right)^{1+n}\right) dt^2 - \left(1 - \left(\frac{r_h}{r}\right)^{1+n}\right)^{-1} dr^2 + r^2 d\Omega^2_{2+n}\,, \label{eq:metric_nD}
\end{equation}
where the horizon radius is given by:
\begin{equation}
 r_h = k_n \frac{1}{M_*} \left(\frac{M}{M_*}\right)^{\frac{1}{1+n}}\,,
\end{equation}
with $M$ the black hole mass and
\begin{equation}
k_n = \left(\frac{8 \Gamma((3+n)/2)}{(2+n)\pi^{(1+n)/2}}\right)^{\frac{1}{1+n}} \,. \label{eq:kn}
\end{equation}

By definition, the field equations in $(4+n)$ dimensions are scale-invariant in empty space. However, the mass scale $M_*$ is embedded in the Schwarzschild metric. In 4 dimensions, $M_* = M_P$, which is fixed and is not modified under transformation (\ref{eq:transformation}). As for the 4 dimensional case, the scaling properties of the metric (\ref{eq:metric_nD}) imply that $r_h \longrightarrow \lambda r_h$, and as a consequence the scaling of the mass reads in $(4+n)$ dimensions:
\begin{equation}
 M \longrightarrow \lambda^{1+n} M\,. \label{eq:xdim_mass}
\end{equation}
Similarly, the scaling properties of gravitational waves are identical to the 4-dimensional case apart from the energy which scales as the mass. The propagation of gravitational waves can be affected by the existence of extra-dimensions since the gravitational waves can propagate into the extra-dimensions, modifying for example the propagation speed.

As a consequence, one can expect deviations from the scaling of the observed gravitational waves with the masses of a black hole merger, which would be a characteristic of the considered extra-dimension model.

We studied here the case of a very simple extra-dimension scenario, but there exist many extra-dimension models, with for example more complicated geometries. In such cases both the emission and propagation of gravitational waves can be modified, altering the scaling properties obtained in this section.

\subsection{Lovelock black holes}

We consider now the case of Lovelock black holes in the 5-dimensional Einstein-Gauss-Bonnet theory \cite{Lovelock:1971yv}. The geometrical part of the effective 4-dimensional action reads:
\begin{equation}
 S_g = \frac{1}{16\pi G}\int d^4x \sqrt{-g} \Bigl(-2\Lambda + R + \alpha (R^2 + R_{\alpha\beta\mu\nu} R^{\alpha\beta\mu\nu} - 4 R_{\mu\nu} R^{\mu\nu}) \Bigr) \,,
\end{equation}
where $\Lambda$ is the cosmological constant and $\alpha$ is a constant. The last term is the so-called Gauss-Bonnet term, which appears as an ultraviolet correction to Einstein's gravity.

Neglecting the cosmological constant, the spherically-symmetric and static solution of the field equations leads to the following metric \cite{Boulware:1985wk,Deser:2005pc,Garraffo:2008hu}:
\begin{equation}
 d\tau^2 = \left(1+\frac{r^2}{4 \alpha} -\frac{r^2}{4 \alpha} \sqrt{1+\frac{8 k_2^2 \alpha M}{M_*^3 r^4}} \right) dt^2 - \left(1+\frac{r^2}{4 \alpha} -\frac{r^2}{4\alpha} \sqrt{1+\frac{8 k_2^2 \alpha M}{M_*^3 r^4}} \right)^{-1} dr^2 - r^2 d\Omega^2_3\,,
\end{equation}
where $k_2$ is given in Eq.~(\ref{eq:kn}) and $M_*$ is the fundamental mass scale in 5 dimensions. This metric reduces to the Schwarzschild metric in 5 dimensions when $\alpha \approx 0$ or $r^2 \gg |\alpha|$. The horizon is located at $r_h=\sqrt{k_2^2 M M_*^{-3} - 2 \alpha}$, hence the condition $M > 3\pi \alpha M_*^3 / k_2^2$. An interesting feature of this metric is that there is no naked singularity at the center.

For this metric to be invariant, first the mass has to follow the scaling relation $M \longrightarrow \lambda^2 M$, which is a particular case of Eq.~\ref{eq:xdim_mass}, and second $\alpha \longrightarrow \lambda^2 \alpha$. However, since $\alpha$ is the fundamental coupling of the Gauss-Bonnet term, it is fixed in the theory and cannot be rescaled. At long distance, since the terms involving $\alpha$ vanish, the gravitational waves emitted by the black hole will have the same properties as in the simple 5-dimension extension of Einstein's gravity presented in Section~\ref{sec:xdim}. At short distance, the Schwarzschild metric is also retrieved, but the effective mass is modified. At intermediate distance, the terms proportional to $r^2$ can dominate. As a consequence, the emitted gravitational waves will be similar to the ones emitted by black hole mergers in Einstein's gravity at low frequencies, with the difference that the masses will scale as $\lambda^2$, but at high frequencies such scaling relations will not hold anymore.

\subsection{Black holes in scalar-tensor gravity}

A generic class of theories is based on the existence of a scalar field coupled to geometrical tensors. For black holes, there exists a ``no-hair conjecture'' \cite{Israel:1967wq,Israel:1967za,Carter:1971zc} which states that black hole solutions of the EFEs are completely characterized by only three parameters, the mass, the electric charge, and the angular momentum. This conjecture generally does not apply to scalar-tensor gravity scenarios, in which the black holes can develop ``scalar hair'' \cite{Sotiriou:2014pfa,Antoniou:2017hxj}.

We consider here the example of black holes in the Einstein-Scalar-Gauss-Bonnet theory \cite{Antoniou:2017acq}, in which the geometrical part of the action reads
\begin{equation}
S_g=\frac{1}{16\pi G}\int{d^4x \sqrt{-g}\left[R-\frac{1}{2}\,\partial_{\mu}\varphi\,\partial^{\mu}\varphi+f(\varphi)R^2_{GB}\right]}\,,
\end{equation}
where $f(\varphi)$ is a generic function of the scalar field $\varphi$ and
\begin{equation}
R^2_{GB}=R_{\mu\nu\rho\sigma}R^{\mu\nu\rho\sigma}-4R_{\mu\nu}R^{\mu\nu}+R^2
\end{equation}
is the quadratic Gauss-Bonnet term. In empty space, the field equations can be obtained from the action \cite{Antoniou:2017acq}:
\begin{align}
&R_{\mu\nu} - \frac12 g_{\mu\nu} R = T^\varphi_{\mu\nu}\,,\\
&\square\varphi+f'(\varphi) R^2_{GB}=0\,,
\end{align}
with 
\begin{equation}
T^{\rm vac}_{\mu\nu}= -\frac{1}{4}g_{\mu\nu}\partial_{\rho}\varphi\partial^{\rho}\varphi+\frac{1}{2}\partial_{\mu}\varphi\partial_{\nu}\varphi - 
\frac{1}{2}(g_{\rho\mu}g_{\lambda\nu}+g_{\lambda\mu}g_{\rho\nu})
\eta^{\kappa\lambda\alpha\beta}\eta^{\rho\gamma\sigma\tau}
R_{\sigma\tau\alpha\beta}
\nabla_{\gamma}\partial_{\kappa}f(\varphi)\,,
\end{equation}
and $\eta^{\rho\gamma\sigma\tau}=\epsilon^{\rho\gamma\sigma\tau}/\sqrt{-g}$.

A scaling $T^{\rm vac}_{\mu\nu} \longrightarrow \lambda^{-2} T^{\rm vac}_{\mu\nu}$ would imply for the scalar field $\varphi \longrightarrow \varphi$ and $f(\varphi) \longrightarrow \lambda^2 f(\varphi)$. These transformations are obviously self-contradictory since $f(\varphi)$ does not vary when $\varphi$ is constant. However, since the last term of $T^{\rm vac}_{\mu\nu}$ is multiplied by $R_{\sigma\tau\alpha\beta}$, in the case of a weak field it will be negligible and will in particular not affect the gravitational waves.

We will now check whether the geometry around black holes can conserve its scaling properties in spite of the presence of the scalar field. In a spherically-symmetric and static metric such as:
\begin{equation}
d\tau^2=A(r){dt}^2-B(r){dr}^2-r^2({d\theta}^2-\sin^2\theta\,{d\phi}^2)\,,
\end{equation}
the EFEs connect $A(r)$ and $B(r)$ to $\phi(r)$. Solving the set of field equations in such a metric is a complicated task, but it is interesting to consider the asymptotic solution at infinity. It reads at second order in $1/r$:
\begin{eqnarray}
A(r)&=& 1-\frac{2GM}{r} \,,\nonumber\\
B(r) &=& \left(1-\frac{2GM}{r}-G^2\frac{16 M^2-P^2}{4r^2}\right)^{-1} \,,\\
\varphi(r)&=&\; \varphi_{\infty}+\frac{GP}{r}+\frac{G^2MP}{r^2} \,,\nonumber
\end{eqnarray}
where $M$ is the effective mass and $P$ is the scalar charge. The value of the scalar field at infinity $\varphi_{\infty}$ does not play any role as long as $f(\varphi_{\infty})=0$. As a consequence, at infinity the Schwarzschild metric is retrieved at first order in $r^{-1}$. The next orders can be written in terms of $P$, so that the scalar charge appears as a new parameter (a ``hair'') to describe black holes in tensor-scalar gravity. It is interesting to compare this expression to the case of the charged black hole in Einstein's gravity, for which $A(r)=B^{-1}(r)=1-\frac{2GM}{r}+\frac{Q^2 G}{4\pi\epsilon_0 r^2}$. As for the electric charge $Q$, the scaling rule associated to scalar charge is therefore the same as for the mass $M$: $P \longrightarrow \lambda P$. This scalar charge is however generally not independent of the mass $M$ and is related to the definition of the $f(\varphi)$ function \cite{Antoniou:2017hxj}.

Concerning the gravitational waves themselves, the ones emitted in the early stages of a black hole merger will be very similar to the ones in Einstein's gravity, and their propagation will be unaffected in weak gravitational fields. However, when the black holes get closer, deviations due to the presence of a scalar charge will appear, and the scaling properties will be changed, unless there exists a regime in which the scalar charge is proportional to the mass.

\subsection{Black holes in other modified gravity models}

There exist many other possibilities to alter Einstein's gravity, we consider here one last case, namely the non-relativistic Ho\v{r}ava-Lifshitz four-dimensional theory of gravity \cite{Horava:2008ih,Horava:2009if}, which is a renormalizable gravity theory in four dimensions which reduces to Einstein's gravity in IR but with improved UV behaviours. We will not consider here the gory detail, but study the case of the asymptotically flat, spherically-symmetric and static case, for which the metric is \cite{Kehagias:2009is,Park:2009zra}:
\begin{equation}
 d\tau^2 = f(r) dt^2 - \frac{dr^2}{f(r)} - r^2 (d\theta^2 + \sin^2\theta d\phi^2)\,,
\end{equation}
where
\begin{equation}
 f(r) = 1 + \omega r^2 - \sqrt{r (\omega^2 r^3 + 4 \omega M)}\,,
\end{equation}
with $\omega$ a fundamental constant of the theory and $M$ a mass parameter. For $r \gg (M/\omega)^{1/3}$, the Schwarzschild metric is retrieved. There are two horizons located at radii:
\begin{equation}
 r_\pm = M \left( 1 \pm \sqrt{1-\frac{1}{2 \omega M^2}}\right)\,.
\end{equation}
Therefore, if $\omega M^2 \ge 1/2$, there is no naked singularity at the center, where the geometry is flat. Once the fundamental parameters are fixed, only $M$ can be changed. For $f(r)$ to be invariant under the transformation $r \longrightarrow \lambda r$, the necessary transformation laws are $\omega \longrightarrow \lambda^{-2} \omega$ and $M \longrightarrow \lambda M$. The scaling of the mass is therefore the same as in Einstein's gravity. Nevertheless, because $\omega$ is a fixed parameter, even if the gravitational waves emitted during a merger of black holes are similar to the ones of Einstein's gravity at long distances and low frequencies, the scaling properties will become invalid at short distances and high frequencies.

\section{Rescaling of simulations of black hole mergers}
\label{sec:simulations}

We now discuss simulations of mergers of binary black holes and emission of gravitational waves, and their limitations in the case where the rescaling properties are not valid. Such simulations are computationally intensive, and each of our simulations necessitated about one week on a 64-processor server. The evolution of binary black holes includes different phases. The first one, which is the longest, is the inspiral phase described using post-Newtonian techniques. The second one is the plunge and merger phase, which can be only described by numerical relativity. The last one is the ringdown phase described by perturbation methods. 

To produce some examples of simulations and solve the Einstein equations, we use the {\tt Einstein Toolkit} code \cite{Thornburg:2003sf, Schnetter:2003rb, Schnetter:2007rb, Loffler:2011ay}. The initial data are generated via the {\tt TwoPunctures} routine \cite{Dreyer:2002mx} for a merger with a near circular orbit. We consider only this case since for a binary orbit with a given eccentricity $e$, the emission of gravitational waves leads to a decrease in eccentricity. The evolution is performed using the BSSN formulation \cite{Baumgarte:2010ndz} via the {\tt McLachlan} routine \cite{Brown:2008sb}. To compute the properties of the emitted gravitational waves, we use the Newman-Penrose formalism \cite{Baumgarte:2010ndz} where the Weyl scalar $\psi_4$ and the GW polarization amplitudes $h_{+, \times}$ are related by: 
\begin{equation}
\ddot{h}_+-i\ddot{h}_{\times}=\psi_4=\sum_{l=2}^{\infty}\sum_{m=-l}^{l} \psi_4^{lm}(t,r){}_{-2}Y_{lm}(\theta, \phi) \,.
\end{equation}
This corresponds to the decomposition of $\psi_4$ into $s=-2$ spin-weighted spherical harmonics. The dominant modes for the gravitational wave strain $h$ are the $l=m=2$ modes: $h^{22}_{+,\times}$. Considering the scaling relations, since $h_{+, \times}$ are scale invariant, we have $\psi_4 \longrightarrow \lambda^{-2} \psi_4$. 

\begin{figure}[!t]
\begin{center}
\includegraphics[width=8.5cm]{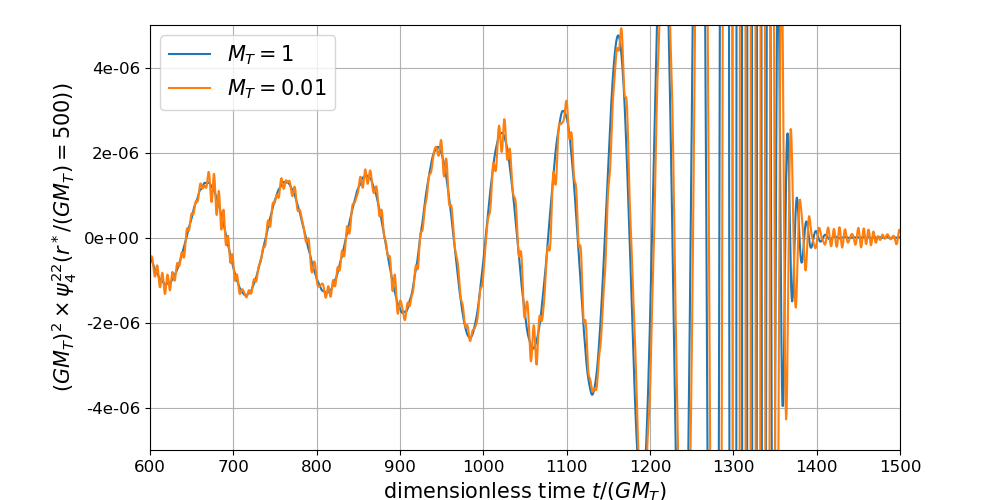}\includegraphics[width=8.5cm]{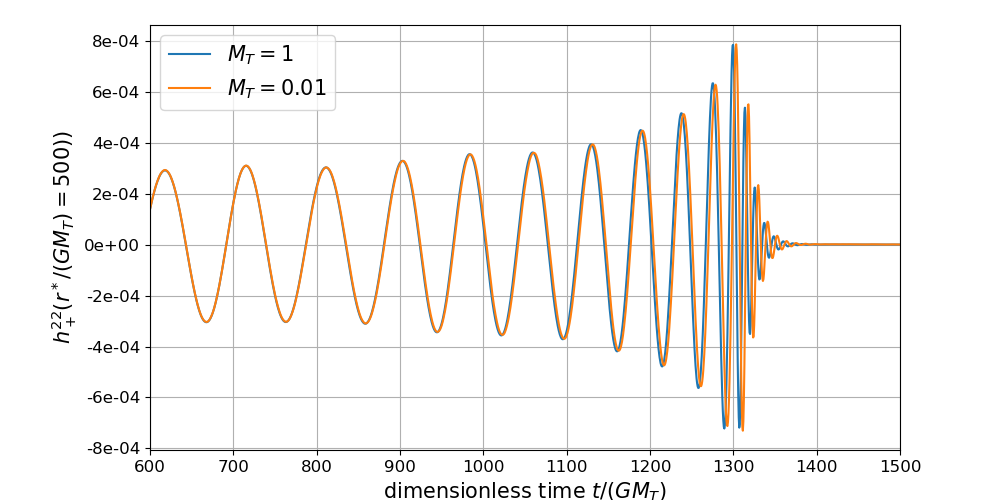}
\end{center}
\caption{Real part of the dimensionless Weyl scalar $\psi_4^{22}$ (left) and GW strain signal $h_+^{22}$ (right) as a function of the dimensionless time, corresponding to the mergers of two Schwarzschild BHs for a total mass $M_T$ of $1$ (blue) and for a total mass $M_T$ of $0.01$ (orange). \label{fig:mergers_S0}} 
\end{figure}

The scaling relations have been used for binary black holes since a long time \cite{Pretorius:2005gq,Centrella:2010mx}, and the results are usually presented in the $c=M_\odot=G=1$ units, where the masses are adimensioned. Setting $G=1$ corresponds in our convention to considering that the distance, time and mass units are equal, or in terms of scaling that they all scale as $\lambda$, which is correct in Einstein's gravity. We first perform two simulations of Schwarzschild BH mergers. The first one is the merger of two BHs of masses equal to 0.5, which can be taken as the reference simulation. The second one is a similar simulation with two BHs of mass 0.005, corresponding to a scale factor of $\lambda=0.01$ and a total mass of 0.01, and for which the code is not optimized since it runs outside its natural units in which the total mass is equal to 1. Such a case would appear when non-standard black holes with different scales are considered, since one scale would need to be chosen to adjust the simulation parameters, and the other scale may be sufficiently different to generate numerical instabilities. To compare both simulations, all the initial quantities are rescaled using the relations given in the Section~\ref{sec:scaling} (e.g. mass of the BHs, initial angular momenta of the two BHs, ...). In Fig.~\ref{fig:mergers_S0}, we show the real part of the Weyl tensor and the GW strain for $l=m=2$, as a function of time. To fulfil the scale invariance, the time and Weyl tensor have been adimensioned using the total initial mass, as given in the axis labels. As expected the two simulations give similar results in these scale invariant parameter planes. The main difference is a high-frequency numerical noise, which comes from interpolation procedures between the different path grids. In order to reduce this noise, it is necessary to finely adjust the strength of the Kreiss-Oliger artificial dissipation term \cite{kreiss-oliger} or change the finite differential order, but this requires re-running simulations with adjusted parameters until the correct precision is reached.  On the contrary, the reference simulation does not show any instability.

\begin{figure}[!t]
\begin{center}
\includegraphics[width=8.5cm]{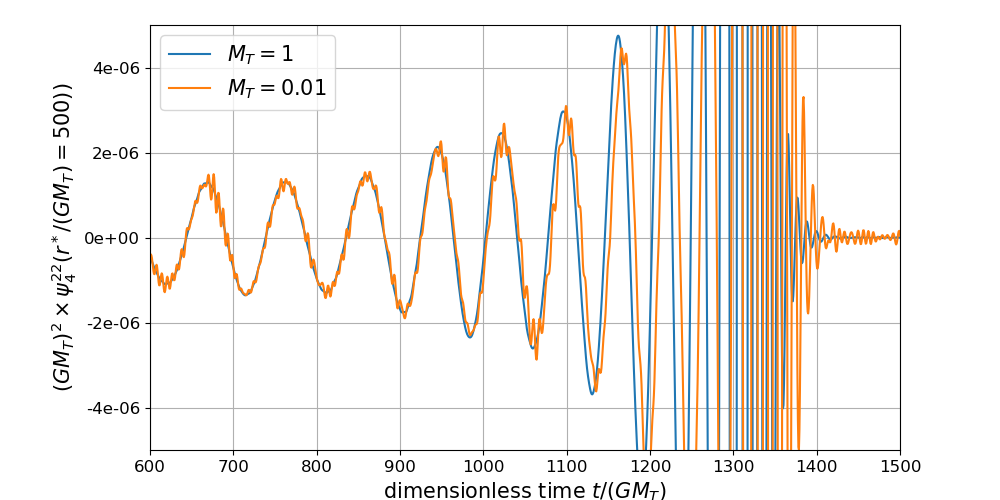}\includegraphics[width=8.5cm]{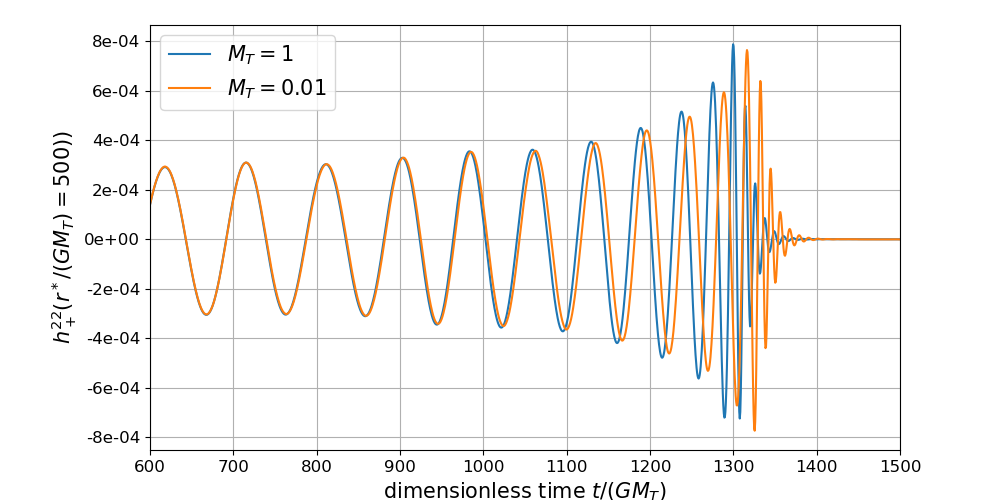}
\end{center}
\caption{Same as Fig.~\ref{fig:mergers_S0} for two spinning BHs of reduced spin parameters $a^*=\pm 0.5$.\label{fig:mergers_S5}}
\end{figure}

Figure \ref{fig:mergers_S5} shows $\psi_4^{22}$ and $h_+^{22}$ for a similar merger, for two Kerr BHs with reduced spin parameters $a^*=\pm 0.5$, as a function of time. Similarly to Fig.~\ref{fig:mergers_S0}, we made two simulations of BH mergers with total masses of $1$ and $0.01$. We see that the numerical noise has in the case of Kerr BHs a larger impact which leads to a shift at later times. This can only be reduced using a very finely adjusted Kreiss-Oliger dissipation strength and/or a change of finite differential method, after which it is mandatory to run many simulations to reach a correct precision.

As discussed in the previous sections, when comparing the spectra of the observed gravitational waves of numerous black hole mergers, a breaking of the scaling relations of Einstein's gravity would be an unambiguous sign of non-standard gravity. The simulations that we have produced show that if there exists a scale in the studied non-standard black hole models which differs by two orders of magnitude from the main scale (set in general by the mass), numerical instabilities can occurs, in particular at high frequencies. This problem already exists when simulating mergers of two black holes with very different masses (even if, when one mass is negligible, a perturbative approach may be possible). In case the gravitational waves of binary black hole mergers reveal a breaking of the scaling relations, it will be necessary to perform simulations in non-standard gravities in order to interpret the results, and to generate templates for gravitational wave experiments. It will then be important to pay attention to the fact that numerical instabilities induced by the presence of several scales may generate a spurious breaking of scaling properties, which may misleadingly be interpreted as non-standard physics. Nevertheless, as we have seen, in non-standard models the new parameters can also follow scaling rules. If they are unknown on the theory side, the scaling relations can be used to reduce the number of simulations to be performed to generate a library of templates spanning the model parameters, and to rescale the resulting templates by rescaling also the fundamental parameters.

\section{Conclusions and perspectives}
\label{sec:conclusion}

In this article, we have studied scaling transformations for black hole mergers, from primordial to supermassive black holes. We have verified that these rescaling relations are not only valid for Schwarzschild and Kerr spinning black holes, but also for Kerr-Newman charged spinning black holes, and that they extend to both the gravitational and electromagnetic waves emitted by such mergers. This opens the way to a rescaling of numerical simulations for the most general cases of black holes, and in particular for primordial black holes which are more likely to have extremal spins and large electromagnetic charges, thus reducing the number of parameters to be varied for a full coverage of the parameter space of black hole mergers. However, primordial black holes have specificities which can lead to a breaking of the scaling relations. We have therefore studied the limitations of the rescaling of primordial black holes: first, in the early Universe, the cosmological expansion can prevent the merger because it increases the distances faster than the merger decreases them. Then, for light primordial black holes, Hawking evaporation can be faster than the merger. Finally, we have considered non-standard black holes models, such as the Morris-Thorne wormhole model, the loop quantum gravity inspired Schwarzschild  hole model, and several cases of modified gravity black hole models, which can have different scaling properties.

The scaling relations for black hole mergers and their emission of gravitation and electromagnetic waves are of utmost importance, because they emerge from the invariance of general relativity and electromagnetism in empty space, and can allow to speed up numerical simulations of numerous templates. As a consequence, with the multiplication of observations of GWs from BH mergers, it will be important to compare the observed spectra of binary black holes with different masses, because deviations from the scaling relations will be a clear sign of the presence of non-standard phenomena and may lead to the discovery of new gravitational physics.

\bibliographystyle{apsrev4-1}
\bibliography{biblio}

\end{document}